# Modified U-Net (mU-Net) with Incorporation of Object-Dependent High Level Features for Improved Liver and Liver-Tumor Segmentation in CT Images

Hyunseok Seo, Charles Huang, Maxime Bassenne, Ruoxiu Xiao, and Lei Xing

*Abstract*—Segmentation of livers and liver tumors is one of the most important steps in radiation therapy of hepatocellular carcinoma. The segmentation task is often done manually, making it tedious, labor intensive, and subject to intra-/inter- operator variations. While various algorithms for delineating organ-at-risks (OARs) and tumor targets have been proposed, automatic segmentation of livers and liver tumors remains intractable due to their low tissue contrast with respect to the surrounding organs and their deformable shape in CT images. The U-Net has gained increasing popularity recently for image analysis tasks and has shown promising results. Conventional U-Net architectures, however, suffer from three major drawbacks. First, skip connections allow for the duplicated transfer of low resolution information in feature maps to improve efficiency in learning, but this often leads to blurring of extracted image features. Secondly, high level features extracted by the network often do not contain enough high resolution edge information of the input, leading to greater uncertainty where high resolution edge dominantly affects the network's decisions such as liver and liver-tumor segmentation. Thirdly, it is generally difficult to optimize the number of pooling operations in order to extract high level global features, since the number of pooling operations used depends on the object size. To cope with these problems, we added a residual path with deconvolution and activation operations to the skip connection of the U-Net to avoid duplication of low resolution information of features. In the case of small object inputs, features in the skip connection are not incorporated with features in the residual path. Furthermore, the proposed architecture has additional convolution layers in the skip connection in order to extract high level global features of small object inputs as well as high level features of high resolution edge information of large object inputs. Efficacy of the modified U-Net (mU-Net) was demonstrated using the public dataset of Liver tumor segmentation (LiTS) challenge 2017. For liver-tumor segmentation, Dice similarity coefficient (DSC) of 89.72 %, volume of error (VOE) of 21.93 %, and relative volume difference (RVD) of -0.49 % were obtained. For liver segmentation, DSC of 98.51 %, VOE of 3.07 %, and RVD of 0.26 % were calculated. For the public 3D Image Reconstruction for Comparison of Algorithm Database (3Dircadb), DSCs were 96.01 % for the liver and 68.14 % for liver-tumor segmentations, respectively. The proposed mU-Net outperformed existing state-of-art networks.

## I. Introduction

According to the World Health Organization (WHO), liver cancer is one of the five most common causes of cancer-induced deaths in 2018 [1]. Liver and tumor target segmentation represent an important step in successful liver radiation therapy and other interventional procedures. Liver tumors have a deformable shape and high variability of location, as well as poor contrast with respect to the surrounding tissues in CT images. Thus, in current practice, segmentation is predominantly done manually, which is time-consuming and suffers from inter- or intra-operator variations [2]. Many algorithms for automatic segmentation of livers and liver tumors have been studied, such as atlas-based models [3], graphical models [4, 5], and deformable models [6, 7]. These model-based methods can offer good quality of segmentation results, but often involves the use of some parametric steps, which are patient specific and restrict the models from being more commonly used. The learning-based model [8] has been proposed for automatic segmentation results based on careful feature engineering, but it is not stable enough to deal with all clinical scenarios due to its high sensitivity to constructed features. Recently, inspired by the tremendous success of deep learning in classification [9, 10], many studies on deep

This work was partially supported by NIH/NCI (1R01 CA176553), Varian Medical Systems, a gift fund from Huiyihuiying Medical Co, and a Faculty Research Award from Google Inc.

H. Seo, M. Bassenne, and R. Xiao are with the Laboratory of Artificial Intelligence in Medicine and Biomedical Physics, Medical Physics Division, Department of Radiation Oncology, School of Medicine, Stanford University, Stanford, CA, 94305, USA (e-mail: hsseo@stanford.edu; bassenne@stanford.edu; xiaoruoxiu@ustb.edu.cn).

C. Huang is with the Department of Bioengineering, School of Engineering and Medicine, Stanford University, Stanford, CA, 94305, USA (e-mail: chh105@stanford.edu).

L. Xing is with the Laboratory of Artificial Intelligence in Medicine and Biomedical Physics, Medical Physics Division, Department of Radiation Oncology, School of Medicine and with the Department of Electrical Engineering, School of Engineering, Stanford University, Stanford, CA, 94305, USA (e-mail: lei@stanford.edu).

learning in classification [9, 10], many studies on deep learning for segmentation have emerged and have been applied for liver [11-14], head and neck [15, 16], prostate [17], and brain [18, 19] segmentations, respectively. Most of these methods are based on the convolution neural networks (CNNs). In particular, the U-Net has shown greatest performance to date. In general, the results of the segmentation depend on the boundary of the input object. U-Net structures have improved segmentation performance by incorporating high-resolution low-level features into the decoding part of the network. Most recent deep learning-based segmentation works use this skip connection to transfer high resolution information across the network. However, one of the drawbacks of the skip connection is the duplication of low resolution contents [20]. One other drawback is that high resolution edge information of the input is not sufficiently represented in high level feature maps extracted by the network. Further, it is generally difficult to optimize the number of pooling operations to extract high level global features. For example, to maintain the context information of the small object, the number of pooling operations used should be less than the number of pooling operations used for the large object due to resolution loss after pooling.

This paper proposes an object-dependent up sampling and redesigns the residual path and the skip connection to overcome the limitations of the conventional U-Net. The modified U-Net (mU-Net) adaptively incorporates features in the residual path into features in the skip connection, and enables (1) to prevent duplication of low resolution information of features; (2) to extract higher level features of high resolution edge information for large objects; and (3) to extract higher level global features for small objects by using the optimal number of pooling operations. In comparison to the conventional U-Net, the mU-Net can more effectively handle edge information and morphologic information of the objects. In the rest of this paper, we provide a detailed mathematical description of the mU-Net and report results from its application to relevant validation studies.

## II. METHODS

### A. Backgrounds with mathematical preliminaries

Both CNNs and U-Nets have a convolution layer which is composed of two essential operations of convolution and activation combined with pooling or up pooling. The convolution layer generates various features in the spatial domain and can also greatly reduce the amount of computation complexity by sharing the kernel coefficients for one feature map. Each layer has multiple convolution kernels. The activation has a similar role to the adaptive filter with respect to the sign of each pixel value of the input generated after convolution, which can impose a non-linearity to the network. The output of the convolution layer (convolution + activation) and pooling in a fully convolutional network (FCN), $c^i = \{c_1^i, c_2^i, \cdots, c_{K_i}^i\}$, calculated from convolution with kernel of $w^i = \{w_1^i, w_2^i, \cdots, w_{K_i}^i\}$, bias value of $b^i = \{b_1^i, b_2^i, \cdots, b_{K_i}^i\}$, and activation in the $i^{th}$ layer is defined as follows,

$$c_j^i = f(w_j^i \otimes \mathcal{D}(c^{i-1}) + b_j^i) \in c^i, 1 \leq j \leq K_i,$$
$$c^i \equiv C^i(\mathcal{D}(c^{i-1}); \theta^i), 1 \leq i \leq l, \quad (1)$$

where $f(\cdot)$ is the activation of $\cdot$, $\otimes$ is the convolution operator, $\mathcal{D}(\cdot)$ is the pooling or down sampling of $(\cdot)$, and $\theta^i$ is the set of all parameters of the $i^{th}$ layer. The pooling operation makes the network handle a large receptive field during global feature extraction. Translational invariance from pooling also significantly reduces the number of parameters. In this study, parametric rectifier linear unit (PReLU) was applied for the activation and is defined as follows [21],

$$f(x) \equiv \max(x, 0) + \alpha \min(0, x), \quad (2)$$

where $\alpha$ is a parameter, and its range is $0 < \alpha < 1$. Then, the final output signal, $C(I; \theta)$, of the feed forward network can be described as follows,

$$(I; \theta) = \tilde{C}^{2l-1}(\cdots \tilde{C}^{l+1}(C^l(\cdots C^2(C^1(I; \theta^1); \theta^2) \cdots; \theta^l); \theta^{l+1}) \cdots; \theta^{2l-1}), \quad (3)$$

where $I$ is the input signal which is equal to $\mathcal{D}(c^0)$, the parameter set of $\theta = \{\theta^1, \theta^2, \cdots, \theta^{2l-1}\}$, and $\tilde{C}^r$ that is the output of the convolution layer in the decoding part is defined as follows,

$$\tilde{c}_q^r = f(w_q^r \otimes \mathcal{U}(\tilde{c}^{r-1}) + b_q^r) \in \tilde{c}^r, 1 \leq q \leq N_r,$$
$$\tilde{c}^r \equiv \tilde{C}^r(\mathcal{U}(\tilde{c}^{r-1}); \theta^r), l+1 \leq r \leq 2l-1. \quad (4)$$

In Eq. (4), $\mathcal{U}(\cdot)$ is the up pooling or up sampling of $(\cdot)$ to recover original matrix size, $w^r = \{w_1^r, w_2^r, \cdots, w_{N_r}^r\}$ is the convolution kernel for the decoding part, $b^r = \{b_1^r, b_2^r, \cdots, b_{N_r}^r\}$ is the bias value for the decoding part in the $r^{th}$ layer, and $\tilde{c}^{r-1}$ is equal to $c^{r-1}$ only if $r = l + 1$. The learning process means that the parameter set of $\theta$ is iteratively updated by backpropagation based on the gradient descent algorithm [22, 23] to minimize a value of the pre-defined loss function as follows,

$$\operatorname{argmin}_\theta G(C(I; \theta); O), \quad (5)$$

where $O$ is the desired output or labeled signal, and $G(\cdot; \hat{\cdot})$ is the predefined loss function that measures the error between $(\cdot)$ and $(\hat{\cdot})$. This type of CNN network produces good results, however, there is loss of the spatial information to pooling. The U-Net [24] overcomes this drawback using skip connections which connects high-resolution low-level features to the decoding part of the network, such as the $1^{st}$ and $2l-1^{th}$ layers, $2^{nd}$ and $2l-2^{th}$ layers, $\cdots$, and $l-1^{th}$ and $l+1^{th}$ layers. So, the U-Net has become the most popular base network widely used in biomedical image segmentation [25-27].

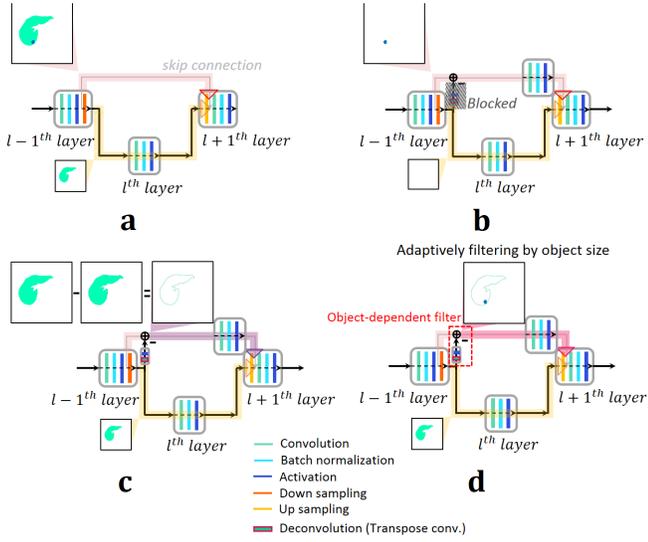

Fig. 1. Schematic diagram of (a) conventional U-Net and (b-d) the proposed mU-Nets. (a) In the case of the conventional U-Net, full information of features passes through the skip connection and only the low resolution information is transferred to the next stage. Spatial information of small objects often disappears due to resolution loss after pooling. (b) In the mU-Net case for small objects, higher level global features can be extracted without loss of resolution by pooling. Spatial information of small objects is maintained by blocking the deconvolution path, which allows small object features to pass into the skip connection without being removed by pooling. (c) In the mU-Net case for large objects, feature information in the skip connection is restricted to edge information to avoid duplication of low resolution information. (d) A schematic diagram of the proposed network is shown. The deconvolution and activation in the residual path adaptively incorporate features of the residual path into features of the skip connection depending on the object size.

### B. Limitation of the skip connection in the U-Net and pooling

Segmentation or contouring processes are usually affected by the edge of the object. Despite the skip connection in the conventional U-Net more effectively handling edge information, there are still some drawbacks of the U-Net. First, as already reported by [20], the U-Net architecture duplicates low resolution information (low frequency components) of features. After pooling (down sampling), low resolution information of features pass on to the convolution layer in the next stage. However, these low resolution information of features are transferred by the skip connection of the U-Net as well, as shown in Fig. 1(a). Duplication of low resolution information may then cause smoothing of the object boundary information in the network, which is more serious in the case of fuzzy object boundaries. Another drawback of the U-Net architecture is that it may not sufficiently estimate high level features for high resolution edge information of the input object. The U-Net can use the skip connection to transfer high resolution information. However, unlike low resolution features after pooling, high resolution edge information does not pass through any convolution layers during transfer by the skip connection, as shown in Fig. 1(a). Thus, higher level feature maps learned by the network do not contain enough of the input-object's high resolution edge information. The input signal, $p^{l+1}$, before convolution in $l+1^{th}$ layer at the decoding part in Fig. 1(a) can simply describe the drawbacks of the U-Net as follows,

$$p^{l+1} = c^{l-1} \oplus \mathcal{U}(c^l) = c^{l-1} \oplus \mathcal{U}(C^l(\mathcal{D}(c^{l-1}); \theta^l)). \quad (6)$$

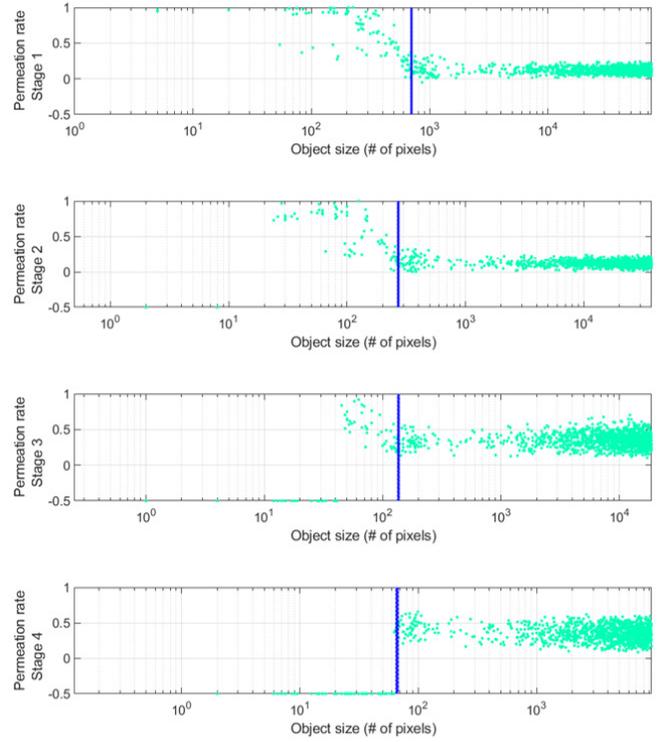

Fig. 2. Permeation rate of the mU-Net with respect to stages. In this work, the stage is related to the matrix size of features. i.e., the same matrix size of features is extracted in the same stage. Blue dashed lines correspond to the size of 28×28 mm² at each stage.

From Eq. (6), $\oplus$ concatenates two signals before and after the operator. The signal, $p^{l+1}$, has duplicated information of $\mathcal{D}(c^{l-1})$, because $c^{l-1}$ already contains its low resolution information of $\mathcal{D}(c^{l-1})$ which is propagated from the further convolution layer of $C^l$. High resolution edge information included only in $c^{l-1}$ do not pass through many convolution layers. In contrast, $\mathcal{D}(c^{l-1})$, which is low resolution information of $c^{l-1}$, passes through more convolution layers, described in Eq. (6), *e.g.*, $C^l$. Thus, in the conventional U-Net, high level features are extracted disproportionately from low resolution information.

### C. Proposed mU-Net architecture

The proposed network uses a residual path to avoid duplication of low resolution feature map information. But, unlike the previous study from [20], the proposed network places the residual path on right after pooling, as shown in Figs. 1(b-d). By doing so, conceptually, high resolution edge information of the feature maps passing through the skip connections are controlled adaptively and are finally combined with additional convolution layers in the skip connection, as shown in Figs. 1(b-d). To assess the performance of this adaptive filter, a permeation rate was defined, as follows,

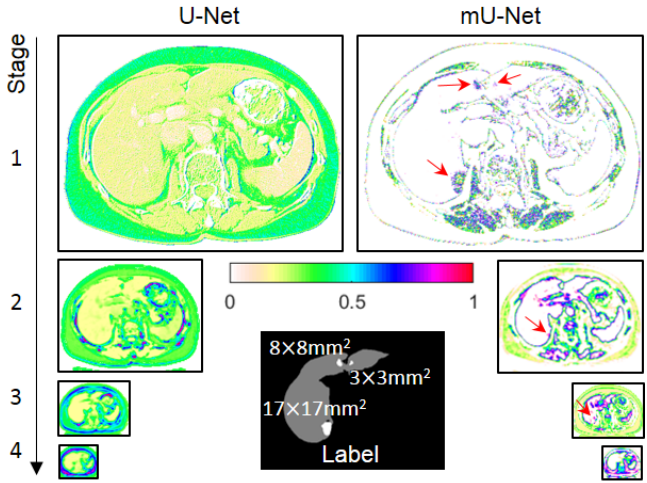

Fig. 3. Feature maps passing through the skip connection of the conventional U-Net (left) and feature maps passing through the skip connection before the additional convolution layer of the mU-Net (right). Red arrows show that, unlike large objects, the features of small objects are preserved in the mU-Net. The tumor sizes are represented in the label image.

$$\text{Permeation rate} = \begin{cases} -0.5, & \text{if } FM_b(x,y) < 0.01, \\ \dfrac{\sum_{x,y \in \text{object}} \text{label}(x,y) FM_a(x,y)}{\sum_{x,y \in \text{object}} \text{label}(x,y) FM_b(x,y)}, & \text{otherwise,} \end{cases} \quad (7)$$

where $FM_a$ is the normalized feature map in the skip connection right after the residual path, $FM_b$ is the normalized feature map right before the residual path, and label means the binary mask of the object (liver or liver tumor). Each normalized feature map has a range of [0, 1]. We set the permeation rate to -0.5 when $FM_b(x,y) < 0.01$, which implies that there are no meaningful features for the skip connection.

A small number of pooling layers might be enough to extract global features of the small object due to loss of spatial information after pooling, as shown in Fig. 1(b). We defined an object whose size is less than 65 pixels (28×28 mm$^2$) at stage 4 as a small object, as the network does not keep any information less than this size at this stage (Fig. 2). More efficient feature extraction in the case of small objects occurs when less information is lost due to pooling. We can achieve this by blocking the features at the deconvolution layer in the residual path to preserve the small object information and by placing more convolution layers in the skip connection to extract higher level features, as shown in Fig. 1(b). This modification allows permeation rates of small object features to remain high at early stages (Fig. 2) maintaining information that might be lost after further poolings. Small object information is eventually lost as the stage increases, which is shown in Fig. 2 with permeation rates of -0.5. To improve the efficiency of feature extraction for large objects, features of a large object in the skip connection should be restricted to edge information, as shown in Fig. 1(c). There is less need to extract the low resolution information as it already propagated to later stages, as described in Fig. 1(a).

In the proposed network (mU-Net), up sampling implemented by a deconvolution layer (transposed convolution and activation) in the residual path and a residual operation at the skip connection adaptively filter out the information based on the object size, as shown in Fig. 1(d). $c_u^{l+1}$ that is a signal after proposed object-dependent up sampling in the residual path, and $c_s^{l+1}$ that is a signal after additional convolution layer in short cut by the skip connection are defined, respectively, as follows,

$$c_{u,v}^{l+1} = f\big(w_{u,v}^{l+1} \otimes^T \mathcal{D}(c^{l-1}) + b_{u,v}^{l+1}\big) \in c_u^{l+1},$$
$$c_u^{l+1} \equiv \check{C}_u^{l+1}(\mathcal{D}(c^{l-1}); \theta_u^{l+1}), \quad (8)$$

$$c_{s,t}^{l+1} = f\big(w_{s,t}^{l+1} \otimes (c^{l-1} - c_u^{l+1}) + b_{s,t}^{l+1}\big) \in c_s^{l+1},$$
$$c_s^{l+1} \equiv C_s^{l+1}(c^{l-1} - c_u^{l+1}; \theta_s^{l+1}), \quad (9)$$

where $\otimes^T$, $w_{u,v}^{l+1}$, $b_{u,v}^{l+1}$, and $\theta_u^{l+1}$ are the transposed convolution, the $v^{th}$ kernel weighting, the $v^{th}$ bias value, and set of all parameters for up sampling process in the residual path that is connected to the skip connection combined with $(l+1)^{th}$ layers, respectively. $w_{s,t}^{l+1}$, $b_{s,t}^{l+1}$, and $\theta_s^{l+1}$ in Eq. (9) are the $t^{th}$ kernel weighting, the $t^{th}$ bias value, and set of all parameters for convolution layer in short cut by the skip connection combined with $(l+1)^{th}$ layers, respectively. Then, the signal, $p_{proposed}^{l+1}$, before the convolution operation in $(l+1)^{th}$ layer at the decoding part in Fig. 1(d) can be described as follows,

$$p_{proposed}^{l+1} = c_s^{l+1} \oplus \mathcal{U}(c^l)$$
$$= C_s^{l+1}(c^{l-1} - c_u^{l+1}; \theta_s^{l+1}) \oplus \mathcal{U}(C^l(\mathcal{D}(c^{l-1}); \theta^l)). \quad (10)$$

From Eq. (10), feature information in the residual path is adaptively filtered by $c^{l-1} - c_u^{l+1}$ unlike the conventional U-Net in Eq. (6). To simplify, let's assume that adaptive up sampling results in an interpolation for features from the large object and an annihilation filtering for features from the small object at the early stage so that permeation rates are high for a small object and low for a large object, as shown in Fig. 2. Then, the effects of the proposed network are as follows,

(Features from a large object)
$$p_{proposed}^{l+1} = C_s^{l+1}(c^{l-1} - c_{low}^{l-1}; \theta_s^{l+1}) \oplus \mathcal{U}(C^l(\mathcal{D}(c^{l-1}); \theta^l)),$$

$$\text{where } c_u^{l+1} = \mathcal{U}(\mathcal{D}(c^{l-1})) \equiv c_{low}^{l-1}. \quad (11)$$

(Features from a small object)
$$p_{proposed}^{l+1} = C_s^{l+1}(c^{l-1}; \theta_s^{l+1}) \oplus \mathcal{U}(C^l(\mathcal{D}(c^{l-1}); \theta^l)),$$

$$\text{where } c_u^{l+1} = 0. \quad (12)$$

Here, feature maps of large objects generated after the residual pass in the skip connection have edge-like information, as shown in Fig. 3. The proposed network preferentially extracts edge information for large objects, which is matched to $c_u^{l+1} = c^{l-1} - c_{low}^{l-1}$ in Eq. (11). In contrast, feature maps of small objects do not suffer the same resolution losses as in conventional U-net architectures (i.e., no loss from pooling and deconvolution) and are better in extracting global features, which are matched to $c_u^{l+1} = c^{l-1}$ in Eq. (12) and shown in Fig.

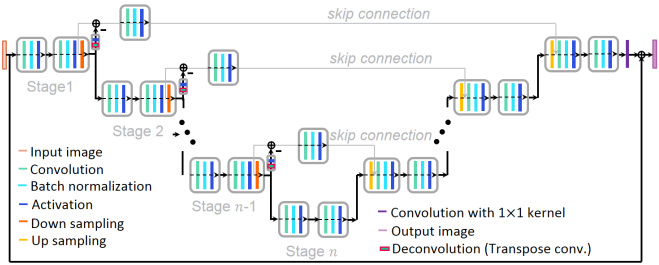

Fig. 4. The proposed network architecture.

3 with red arrows. In addition, both $c^{l-1} - c^{l-1}_{low}$ and $c^{l-1}$ in Eqs. (11) and (12) pass the additional convolution layer of $C_s^{l+1}$ in the skip connection to extract higher level features.

To improve accuracy, the proposed network also uses batch normalization [28], dropout [29], and weight decay [30]. The loss function, $G$, for the proposed network is defined as the mean square error (MSE) between estimated and desired outputs for multi-class segmentation.

### D. Image dataset and data preparation

This study used the public dataset for the liver and liver-tumor segmentation that was obtained from Liver Tumor Segmentation Challenge (LiTS-ISBI2017). The dataset was acquired from 130 abdomen contrast computed tomography (CT) scans. Input size is 512×512, and in-plane resolution has a range from 0.98×0.98 mm² to 0.45×0.45 mm². The number of slices has a range from 75 to 987 with thickness from 5 mm to 0.45 mm. Data from Forty patients were used for training (total 22,500 images) and five patient data (total 2,550 images) were used for validation. The other thirty-five patient data (total 16,125 images) were used for test. More details of the dataset can be found online. The 3D Image Reconstruction for Comparison of Algorithm and DataBase (3Dircadb) was also employed to validate the performance of the proposed mU-Net. Here, input size is 512×512, and in-plane resolution has a range from 0.86×0.86 mm² to 0.56×0.56 mm². The number of slices ranges from 74 to 260 with thickness from 4 mm to 1 mm. In this case, data from fifteen patients were used for training (total 2,295 images), and the other five patients' data (total 525 images) were used for testing. For the proposed method, no preprocessing was performed except scaling of the intensity range from -250 to 250. The URLs for both datasets are provided in the Data Availability section.

### E. Learning parameters and training details

A huge amount of parameters for training were initialized using a truncated normal distribution with mean of 0, standard deviation of 0.05, and constant bias values of 0.1. Then, parameters were updated by the adaptive moment (Adam) algorithm [31] with an adaptive learning rate to improve learning efficiency. The starting value of the learning rate was empirically chosen as 0.0001 to avoid divergence and to improve the speed of convergence, and it was scaled by 0.9 for every 5,000 epochs. A kernel size of 3×3 size for convolution and deconvolution layers in the residual and the skip connection was chosen with a stride of 1 and 2, respectively. Each skip connection has one convolution layer without change of the number of feature maps. The base hyperparameters of the networks that corresponded to the U-Net were consistent with the original U-Net structure [24]. Table I shows the network dimension of the proposed network in detail. Decay of moving average for batch normalization was set to 0.9. Weight decay and probability of dropout for regularization was 0.003 and 0.8, respectively. The proposed method used not a patch, but a full image for input. The stage in Fig. 4 shows the number of pooling operations, which was 5. In this paper, the number of layers are same as the number of activations. The size of each batch was 15 and was selected by considering memory constraints and learning time. The batch was shuffled during every iteration for training. Residual learning [32] was also applied to the proposed method for improving the network performance. All computations for learning were performed on a DGX Station from NVIDIA running Linux operating system with an Intel Xeon E5-2698 v4 2.2 GHz (20-Core) CPU and two of four total Tesla V100 (32 GB memory for each GPU) GPUs. The network architecture was implemented in the well-known deep learning framework, Tensor flow [33].

### F. Performance evaluation

To verify the performance between the ground truth and test results of the proposed network, a total of five objective and common metrics for evaluating segmentation models were utilized. With the labeled data as the ground truth, dice similarity coefficient (DSC) was calculated between binary segmentation masks and is defined as follows,

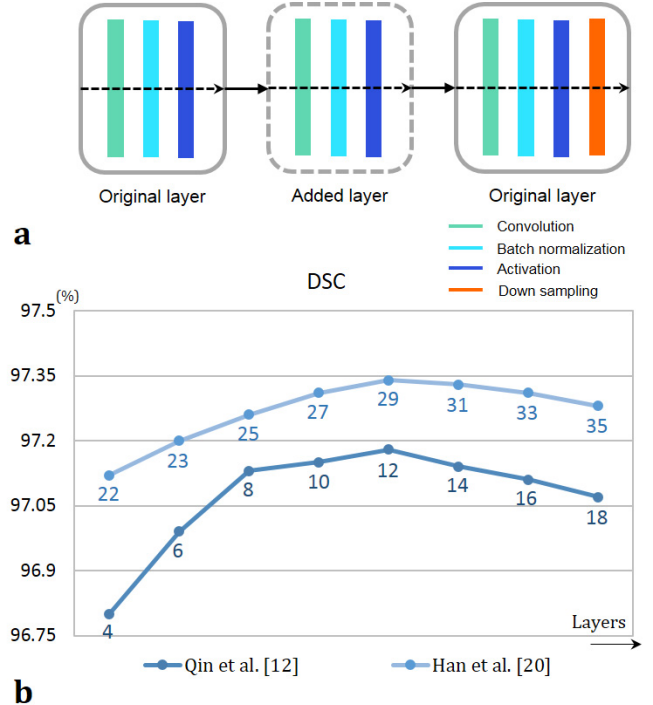

Fig. 5. (a) Adding a new layer in the original ones. (b) The DSC results of liver segmentation in the test cases with respect to the number of layers. The digits below each graph represent the number of layers of each network.

TABLE I
ARCHITECTURE OF THE PROPOSED NETWORK. ALL CONVOLUTIONAL LAYERS INCLUDE THE DROPOUT AND BATCH NORMALIZATION. NOTE THAT "3×3|64" CORRESPONDS TO 3×3 KERNEL GENERATING 64 FEATURES AND TCONV MEANS THE TRANSPOSED CONVOLUTION.

| Stage | Encoding part | Object-dependent up sampling | Skip connection | Decoding part | Fully connected |
|---|---|---|---|---|---|
| 1 | $[3\times3\|64 \quad conv, ReLU]$ $[3\times3\|64 \quad conv, ReLU]$ $[2\times2 \quad \max pool]$ | $[3\times3\|64 \quad tconv, ReLU]$ | $[3\times3\|64 \quad conv, ReLU]$ | $[2\times2 \quad deconv]$ $[3\times3\|64 \quad conv, ReLU]$ $[3\times3\|64 \quad conv, ReLU]$ | $[1\times1 \quad conv]$ |
| 2 | $[3\times3\|128 \quad conv, ReLU]$ $[3\times3\|128 \quad conv, ReLU]$ $[2\times2 \quad \max pool]$ | $[3\times3\|128 \quad tconv, ReLU]$ | $[3\times3\|128 \quad conv, ReLU]$ | $[2\times2 \quad deconv]$ $[3\times3\|128 \quad conv, ReLU]$ $[3\times3\|128 \quad conv, ReLU]$ | - |
| 3 | $[3\times3\|256 \quad conv, ReLU]$ $[3\times3\|256 \quad conv, ReLU]$ $[2\times2 \quad \max pool]$ | $[3\times3\|256 \quad tconv, ReLU]$ | $[3\times3\|256 \quad conv, ReLU]$ | $[2\times2 \quad deconv]$ $[3\times3\|256 \quad conv, ReLU]$ $[3\times3\|256 \quad conv, ReLU]$ | - |
| 4 | $[3\times3\|512 \quad conv, ReLU]$ $[3\times3\|512 \quad conv, ReLU]$ $[2\times2 \quad \max pool]$ | $[3\times3\|512 \quad tconv, ReLU]$ | $[3\times3\|512 \quad conv, ReLU]$ | $[2\times2 \quad deconv]$ $[3\times3\|512 \quad conv, ReLU]$ $[3\times3\|512 \quad conv, ReLU]$ | - |
| 5 | $[3\times3\|1024 \quad conv, ReLU]$ $[3\times3\|1024 \quad conv, ReLU]$ | - | - | - | - |

$$\text{VOE}(A, B) = 1 - \frac{|A\cap B|}{|A\cup B|} \times 100 \ (\%). \tag{14}$$

The relative volume difference (RVD) is the metric for the relative difference of two object volumes and is defined as follows,

$$\text{RVD}(A, B) = \frac{|B|-|A|}{|A|} \times 100 \ (\%). \tag{15}$$

Both VOE and RVD are 0 % when A and B have the same segmentation region. Average symmetric surface distance (ASSD) and maximum symmetric surface distance (MSSD) were also calculated for the distance error [34]. The proposed network was compared to four state-of-art networks based on CNNs and U-Nets, all of which were built to cope with resolution loss of pooling. Two of comparison methods were tested with respect to the number of network parameters as well. To increase the number of parameters, new layers were added in between the original layers in the state-of-art networks designed by Qin *et al*., [12] and Han *et al*., [20], as shown in Fig. 5(a), so that they could have deeper architecture for performance optimization. The number of stage was not changed to avoid loss of context information (i.e., the same number of stages in the original papers). The number of original layers of Men's [35] network, which has a pyramid pooling structures, and Li's [36] network, which has a hybrid DenseU-Net structure with the most superior performance of liver segmentation so far, already much exceeded that of the proposed method, so they were not included in this test about increasing parameters. For expansion to tumor segmentation from Qin's method [12], the output class was increased from three to five (liver boundary, liver, tumor boundary, tumor, and background). All processing for data analysis were implemented using MATLAB (9.4.0.813654, R2018a, The MathWorks Inc., Matrick, MA).

## III. RESULTS

Figure 5(b) shows the DSC performance of liver segmentation with respect to the number of layers (parameters). From the training data set used in this study, the number for the best DSC performance from Qin *et al*. [12] and Han *et al*. [20]

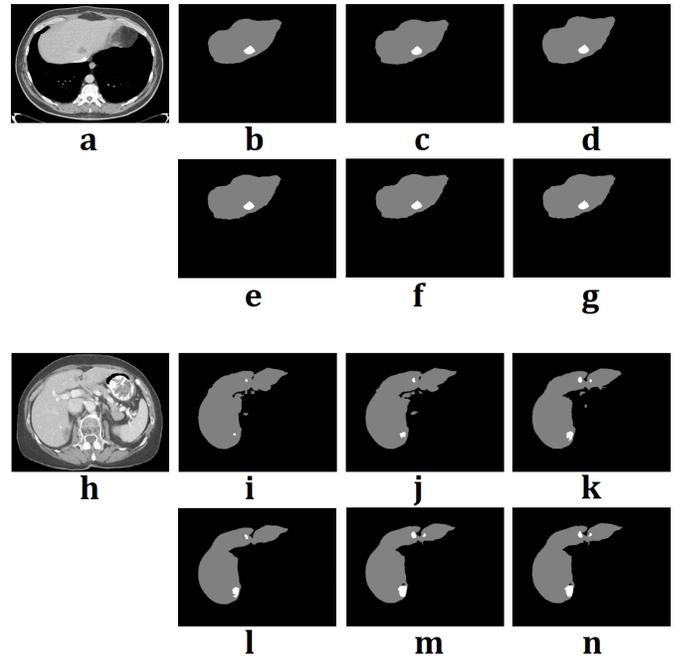

Fig. 6. Target thin slice. Segmentation results from (b) Qin et al. [12], (c) Han et al. [20], (d) Men et al. [35], (e) Li et al. [36], (f) the proposed network, and (g) ground truth. (h) Target thick slice. (i-n) were acquired by the methods corresponded to (b-g), respectively. Gray regions mean liver and white regions mean liver tumor.

was 12 and 29, respectively. With a well-trained model of each network, performances were measured for the segmentation. Figure 6 shows the segmentation results from a thin slice (0.8 mm) and a thick slice (5 mm). Absolute difference maps (error maps) between segmentation results of each method and ground truth are also illustrated in Fig. 7. Figure 8 shows contouring using the results of Fig. 6. For a thin slice, the segmentation and contouring results of the proposed network are the most accurate, however, other methods for comparison offer quite similar results for both liver and liver-tumor segmentation with high accuracy as well. Unlike the results from the thin slice, in the thick slice case, the proposed network has obviously excellent segmentation and contouring results of both liver and liver tumor in comparison to other methods. When verifying the

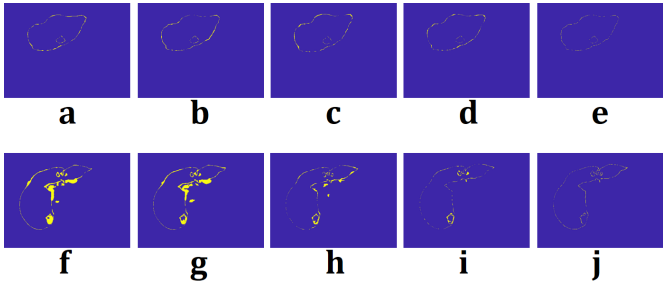

Fig. 7. Absolute difference map between segmentations obtained from (a) Qin et al. [12], (b) Han et al. [20], (c) Men et al. [35], (d) Li et al. [36], (e) the proposed network and ground truth for thin slice. (f-j) were absolute difference map between the methods corresponded to (a-e) and ground truth, respectively, for thick slice. Difference from ground truth is represented with yellow color.

TABLE II
QUANTITATIVE SCORES OF THE LIVER-SEGMENTATION RESULTS (LiTS DATASET). ALL METRIC IS DESCRIBED IN DETAIL IN [34].

|  | Liver | | | | |
|---|---|---|---|---|---|
|  | DSC (%) | VOE (%) | RVD (%) | ASSD (mm) | MSSD (mm) |
| Qin et al. [12] | 97.18 ± 1.22 | 5.81 ± 2.48 | 0.91 ± 0.19 | 1.80 ± 0.55 | 12.48 ± 5.12 |
| Han et al. [20] | 97.36 ± 1.29 | 5.05 ± 2.29 | 0.72 ± 0.14 | 1.81 ± 0.56 | 13.75 ± 5.38 |
| Men et al. [35] | 97.50 ± 1.06 | 3.94 ± 2.17 | 0.45 ± 0.12 | 1.47 ± 0.38 | 10.35 ± 3.78 |
| Li et al. [36] | 98.05 ± **1.01** | 3.31 ± **2.00** | 0.32 ± **0.10** | 1.16 ± **0.35** | 9.17 ± **3.64** |
| mU-Net | **98.51** ± 1.02 | **3.07** ± 2.01 | **0.26** ± **0.10** | **0.92** ± 0.37 | **8.52** ± 3.65 |

TABLE III
QUANTITATIVE SCORES OF THE LIVER-TUMOR-SEGMENTATION RESULTS (LiTS DATASET). ALL METRIC IS DESCRIBED IN DETAIL IN [34].

|  | Liver tumor | | | | |
|---|---|---|---|---|---|
|  | DSC (%) | VOE (%) | RVD (%) | ASSD (mm) | MSSD (mm) |
| Qin et al. [12] | 39.82 ± 24.79 | 61.39 ± 29.04 | -5.34 ± 0.87 | 4.70 ± 1.49 | 13.81 ± 4.72 |
| Han et al. [20] | 55.42 ± 26.37 | 50.73 ± 23.06 | -0.82 ± 0.18 | 1.54 ± 0.43 | 5.99 ± 3.09 |
| Men et al. [35] | 83.14 ± 6.25 | 29.73 ± 16.31 | -0.62 ± 0.14 | 0.96 ± 0.24 | 5.01 ± 1.98 |
| Li et al. [36] | 86.53 ± 5.32 | 24.46 ± 14.43 | -0.53 ± **0.13** | 0.83 ± 0.22 | 4.74 ± 1.97 |
| mU-Net | **89.72** ± **5.07** | **21.93** ± **13.00** | **-0.49** ± 0.15 | **0.78** ± **0.20** | **4.53** ± **1.95** |

results from all slices, 3D volumes of the segmentation results from the thick slice are visualized, as shown in Fig. 9. Volume of liver and liver-tumor segmentation obtained from the proposed network is most similar to that from the ground truth. Figure 10 also shows the overlaid images including liver and liver-tumor segmentation results of the mU–Net and the ground truth, acquired from various test cases. The quantitative analysis from all test cases in Tables II and Table III also show the scores consistent with the results in Figs. (6-10). The proposed mU-Net also attained higher scores than listed in other public databases like 3Dircadb, as listed in Table III and Table IV. Training time/epoch and evaluation time are listed in Table IV as well. The fastest network was Qin's network, because it is based on the CNN without data concatenation from the skip connection and without up sampling. The training time/epoch

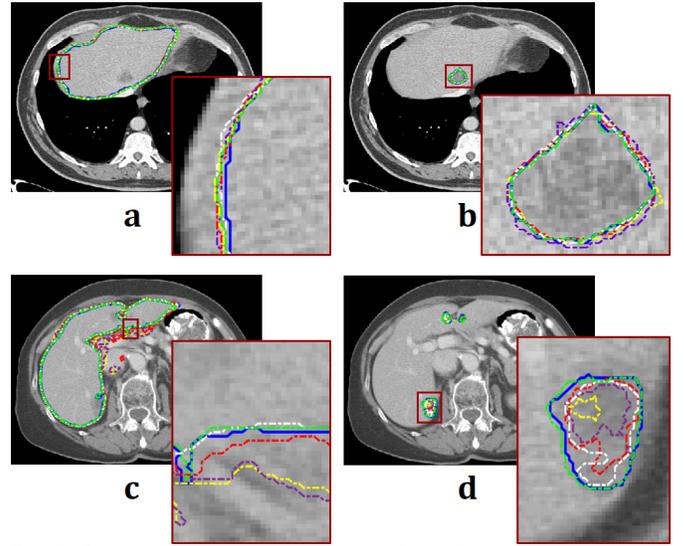

Fig. 8. Contouring results of each method. (a) Liver contouring and (b) liver-tumor contouring from thin slice. (c) and (d) from thick slice correspond to (a) and (b), respectively. Blue, yellow, purple, red, white, and green lines are acquired from ground truth, Qin et al. [12], Han et al. [20], Men et al. [35], Li et al. [36], and proposed network, respectively. Each brown-square region is also magnified.

TABLE IV
QUANTITATIVE SCORES OF THE LIVER-SEGMENTATION RESULTS (3DIRCADB DATASET). ALL METRIC IS DESCRIBED IN DETAIL IN [34].

|  | Liver | | | | |
|---|---|---|---|---|---|
|  | DSC (%) | VOE (%) | RVD (%) | ASSD (mm) | MSSD (mm) |
| Qin et al. [12] | 93.88 ± 1.28 | 12.15 ± 3.62 | 1.06 ± 0.24 | 5.29 ± 1.32 | 13.87 ± 4.82 |
| Han et al. [20] | 94.05 ± 1.20 | 10.94 ± 3.50 | 0.84 ± 0.17 | 4.17 ± 1.20 | 14.34 ± 4.93 |
| Men et al. [35] | 94.32 ± 1.13 | 10.16 ± 3.24 | 0.52 ± 0.15 | 3.84 ± 1.01 | 11.04 ± 4.57 |
| Li et al. [36] | 95.11 ± **1.04** | 9.88 ± 2.83 | 0.39 ± **0.11** | 3.52 ± 0.88 | 9.35 ± 3.95 |
| mU-Net | **96.01** ± 1.08 | **9.73** ± **2.91** | **0.38** ± 0.12 | **3.11** ± **0.84** | **9.20** ± **3.43** |

TABLE V
QUANTITATIVE SCORES OF THE LIVER-TUMOR-SEGMENTATION RESULTS (3DIRCADB DATASET). ALL METRIC IS DESCRIBED IN DETAIL IN [34].

|  | Liver tumor | | | | |
|---|---|---|---|---|---|
|  | DSC (%) | VOE (%) | RVD (%) | ASSD (mm) | MSSD (mm) |
| Qin et al. [12] | 32.66 ± 20.92 | 68.26 ± 24.21 | -10.83 ± 1.42 | 7.49 ± 2.17 | 16.73 ± 6.49 |
| Han et al. [20] | 48.13 ± 18.44 | 54.09 ± 21.71 | -1.11 ± 0.30 | 2.39 ± 0.84 | 7.24 ± 2.54 |
| Men et al. [35] | 64.02 ± 7.18 | 41.37 ± 17.58 | -0.90 ± 0.27 | 1.86 ± 0.60 | 6.42 ± 2.16 |
| Li et al. [36] | 66.47 ± 6.54 | 39.83 ± **13.43** | -0.74 ± **0.18** | 1.71 ± 0.52 | 5.96 ± 2.10 |
| mU-Net | **68.14** ± **6.40** | **36.25** ± 13.82 | **-0.72** ± **0.18** | **1.58** ± **0.51** | **5.91** ± **2.07** |

of Li's networks were the longest, and the evaluation was more than 3 sec.

## IV. DISCUSSION

In this study, we proposed the mU-Net for fully automatic liver and liver-tumor segmentation. In general, deep-learning based segmentation methods estimate local and global features during training by taking advantage of a priori information learned from a large data population. There are a lot of networks for deep learning, and the U-Net has become the most popular network for biomedical image segmentation because it compensates loss of spatial information. However, the U-Net has noteworthy drawbacks for segmentation. One is duplication of low resolution information, and the other is insufficient feature extraction for high resolution edge information. In addition, the number of poolings needed to extract high level global features from large objects may not be appropriate for small objects. Thus, the conventional U-Net cannot effectively use boundary and small object information, despite being able to compensate by transferring high resolution information from the encoding to the decoding sides of the network. In the proposed network, the residual path that prevents the duplication of low resolution information is directly connected to the skip connection with an object-dependent up sampling. This object-based up sampling adaptively incorporates features in the residual path into features in the skip connection. So, in case of the large object, only high resolution edge information of features are transferred on to the skip connection and, in case of the small object, full feature information is transferred. Then, by adding a convolution layer to the skip connection, higher level features can be estimated from high resolution edge information of features from the large object, and higher level global features of the small object can also be extracted. Thus, the proposed network can more efficiently use high resolution edge and small object information, which is related to the segmentation performance of the object.

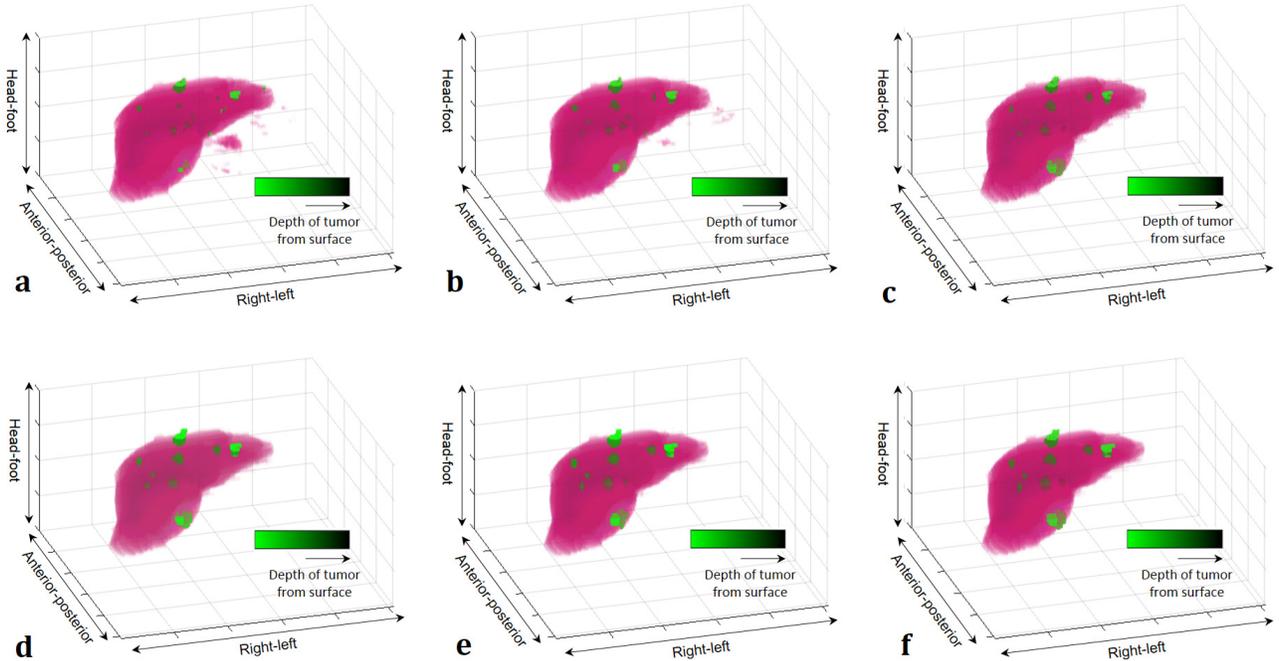

Fig. 9. 3D visualization results of (a) Qin et al. [12], (b) Han et al. [20], (c) Men et al. [35], (d) Li et al. [36], (e) proposed network, and (f) ground truth from segmentation results in Fig. 5. Liver and liver tumor are represented by pink and green color, respectively. Distance of liver tumor from surface is represented by brightness of green color.

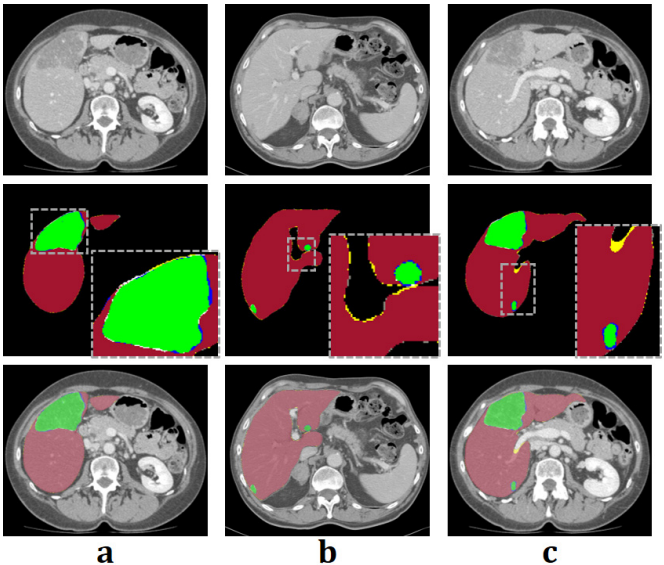

Fig. 10. The example of segmentation results of the mU-Net. First row shows the target slices of (a), (b), and (c). Segmentation results are shown in the second row. Yellow and white regions denote false positive error for liver and liver tumor, respectively. In contrast, gray and blue regions denote false negative error for liver and liver tumor, respectively. Third row represents the images overlaid from first and second row.

TABLE VI
TOTAL NUMBER OF PARAMETERS (LAYERS), TRAINING TIME FOR AN EPOCH, AND TEST TIME OF EACH METHOD.

| | Number of parameters | Training time/epoch (sec) | Evaluation time (sec) |
|---|---|---|---|
| Qin et al. [12] | 1,754,780 (layer: 12) | **97** | **0.84** |
| Han et al. [20] | 3,858,420 (layer: 29) | 2910 | 1.10 |
| Men et al. [35] | 123,559,040 (layer: 121) | 6405 | 1.86 |
| Li et al. [36] | 205,934,600 (layer: 232) | 10970 | 3.25 |
| mU-Net | 4,086,690 (layer: 30) | 3050 | 1.14 |

Partial volume effect in the input image from thick slices, as shown in Fig. 6(h), appears to be severe. Since, there are lots of mixture information in a single slice, boundary of the liver and its tumor is not obvious, as compared to thin slices. So, partial volume effect causes blur to the object boundary. Other networks, which cannot use edge and small object information effectively, are susceptible to unclear boundaries in thick slices. Men *et al*. [35] and Li *et al*. [36] offer more improved results, in comparison to those from Qin *et al*. [12] and Han *et al*. [20], but it is too much of a burden to the network in terms of memory (more than 100 layers). For liver cases, it is relatively easier to make a segmentation mask because it has a more obvious shape and coherent intensity level than those of live tumors, as shown in Figs. 8(a) and 8(c). Thus, the accuracy of segmentation for livers was usually higher than that for liver tumors. Although, liver tumors illustrated in Figs. 8(b) and 8(d) have uncertain boundaries and poor contrast with normal tissues, the proposed network also offers the most accurate segmentation results even if it does not have complex network structure in comparison to pyramid-structure- and DenseNet-based networks. Qin's network [12] offers inaccurate results for the liver-tumor segmentation, because superpixels are inadequate to small and poor contrast areas despite adding an unclear boundary class for classification. The proposed network has few false negative as well as false positive segmentation errors for both livers and tumors, as shown in Fig 10.

Although the proposed network is computationally more intensive, it offers the accurate segmentation results during prediction. The proposed network is particularly beneficial for segmentation of the objects with fuzzy boundary and small targets. The proposed network might be used for the detection of breast cancer in mammography or for cartilage segmentation of musculoskeletal (MSK) images as well as for planning radiation therapy treatments. Often, it is difficult to directly compare performance between different deep learning networks, as the performance of each network is generally affected by the number of layers, training dataset, size of kernels, etc. In this study, when all other hyperparameters (size of the convolution kernels, the number of features in each layer, etc.) are fixed, the performance of networks by Qin *et al*. [12] and Han *et al*. [20] is improved slightly as the number of the convolution layers is increased to 12 and 29, respectively, as shown in Fig. 5(b). Because the structures of these networks are relatively simple, there is room to improve by extracting more features with additional convolution layers. The number of convolution layers in each stage was increased as opposed to adding new stages in order to avoid loss of spatial information by the pooling, as shown in Fig. 5(a). On the other hand, the networks by Men *et al*. [35] and Li *et al*. [36] are already highly complex with many convolution layers, thus all hyperparameters were set to the original values reported in [35] and [36].

The current approach may have some limitations that can be further improved. First, the proposed network was designed for effective training of high resolution edge information, but MSE was applied to the proposed network as the loss function of the network due to its fast computation and multi-class segmentation. The calculation for the derivative of MSE for backpropagation is simple, however, it may not adequately capture structure-similarity information. If a loss function can consider high frequency information with a simple derivative such as soft-dice loss [37], the segmentation performance can be improved. Second, the proposed network also has a common drawback of deep learning, i.e., less generalizability. To achieve the best results, the proposed network should be applied to the same input and output image parameters that were used during training. Finally, due to memory limitations, the current segmentation was performed by slice-by-slice during training. Volume-by-volume training can reveal potential benefits for learning 3D shape features.

## V. CONCLUSION

In conclusion, this paper presents a more robust deep learning network for segmentation, which could offer more accurate results than other networks in both liver and tumor regions even where the boundary is not obvious and the target object is small. By including the residual path and a design of object-dependent up sampling, the proposed network avoids duplication of low resolution information, estimates higher level feature maps that better represent high resolution edge information of larger object inputs, and learns to extract even higher level global features for small object inputs. The proposed method does not require any preprocessing, so it could be generally applied to other organs or other images with poor contrast. It might also be extended to medical images acquired from other imaging modalities such as MRI, PET, or ultrasound.

## DATA AVAILABILITY

LiTS: https://competitions.codalab.org/competitions/17094
3DIRCADb: https://www.ircad.fr/research/3dircadb/